\documentclass[fleqn,10pt]{olplainarticle}

\usepackage{graphicx}
\usepackage{multicol,multirow}
\usepackage{amsmath,amssymb,amsfonts}
\usepackage{mathrsfs}
\usepackage{amsthm}
\usepackage{rotating}
\usepackage{appendix}
\usepackage{ifpdf}
\usepackage[T1]{fontenc}
\usepackage{newtxtext}
\usepackage{newtxmath}
\usepackage{textcomp}
\usepackage{xcolor}
\usepackage{lipsum}
\usepackage{subcaption}
\usepackage[colorlinks,allcolors=blue]{hyperref}

\title{Impact of geophysical fields on Deep Learning-based Lagrangian drift simulations}

\author[1]{Daria Botvynko}
\author[1]{Carlos Granero-Belinchon}
\author[2]{Simon van Gennip}
\author[3]{Abdesslam Benzinou}
\author[1]{Ronan Fablet}

\affil[1]{Lab-STICC, UMR CNRS 6285, IMT Atlantique, Odyssey, INRIA, Brest, France}

\affil[2]{Mercator Ocean International, Toulouse, 31400, France}

\affil[3]{ENIB, Lab-STICC, UMR CNRS 6285, Brrest, France}


\begin{abstract}
We assess the influence of different Eulerian geophysical input fields on Lagrangian drift simulations using DriftNet, a learning-based method designed to simulate Lagrangian drift on the sea surface. Two experiments are conducted: a fully numerical experiment (Benchmark B1) and a real-world drifters-based experiment (Benchmark B2). Both experiments are performed in two regions with different ocean dynamics: North East Pacific and Gulf Stream regions. The performance of DrifNet is evaluated with three different metrics: separation distance between simulated and ground-truth trajectories, the normalized cumulative Lagrangian separation and the autocorrelation of Lagrangian velocities. In both regions, results from B1 show that combining assimilated sea surface currents (SSC) with fully observed sea surface height (SSH) leads to greatest improvement in trajectory simulation. This configuration reduces separation distance by over 50\% and significantly decreases normalized cumulative Lagrangian separation and metrics related to velocities autocorrelation functions compared to the baseline using SSC alone. On the other hand, the inclusion of sea surface temperature (SST) either alone or in combination with SSC generally degrades performance. In B2, using satellite-derived SSH, Ekman and winds velocities improves surface drifters trajectories simulation, particularly in the North East Pacific. While the satellite-derived SST in combination with reanalysis-based SSC configuration leads to better trajectories simulation in the Gulf Stream. Overall, we highlight the added value of combining multiple geophysical fields to improve Lagrangian drift simulation on both numerical and real-world experiments.
\end{abstract}

\begin{document}

\flushbottom
\maketitle
\thispagestyle{empty}

\section{Introduction}
Accurate simulation of Lagrangian drift at the ocean surface is essential for a range of applications, from pollutant transport and tracking of marine debris to ecosystem monitoring or search and rescue \citep{search_and_rescue, pawar2016plastic, van2020physical, lagrangian_fokker_planck_plankton}. Reliable simulations remain challenging, as small errors in the underlying flow fields can rapidly amplify along trajectories \citep{Liu2011Tracking}, reflecting the sensitivity of Lagrangian modeling to multiscale ocean variability. 

Traditionally Lagrangian modeling in the ocean consists in applying advection equation given sea surface currents (SSC) velocity fields \citep{Lange2017, de2020mohid}. While being physically consistent, this type of modeling is strongly dependent on the quality of the modeled SSC, which may be more or less accurate, especially in highly dynamical or coastal regions \citep{glo12}. A wide range of complementary geophysical variables have been shown to improve Lagrangian drift simulations in the ocean. These variables include sea surface height (SSH), surface winds, sea surface temperature (SST), or Stokes drift, with their contribution depending on the studied region and space-time scales. For example, SSC derived from observed SSH  combined with surface winds allowed to improve Lagrangian simulations \citep{liu2014evaluation}. SST improved Machine Learning-driven Lagrangian simulations \citep{scott2012estimates, jenkins2023dnn}. Surface winds have also been shown to be one of the drivers of Lagrangian drift \citep{parn2023effects, zhang2020evaluation}, and the combination of Ekman and geostrophic components of SSC explain a large fraction of observed drifter variance \citep{ralph1999wind, liu2014evaluation, dagestad2019prediction}.

Recent studies have shown that Deep Learning approaches can improve Lagrangian simulations \citep{botvynko2025neural, della2025predicting, trong2025comparative, fajardo2024efficient}. In this study we will only focus on the DriftNet, which has been particularly developed to model Lagrangian drift from geophysical input data and has been shown to outperform state-of-the-art advection-based and Deep Learning-based Lagrangian modeling methods. We build on this framework in order to assess the role of different geophysical fields in Lagrangian trajectories simulations using DriftNet. By training and testing DriftNet with various combinations of SSC, SSH, SST, winds and Ekman current (both from models and observational products), we aim to conclude on their contributions to more accurate Lagrangian drift modeling across chosen metrics. Our analysis focuses on two dynamically distinct regions, the North East Pacific and the Gulf Stream, additionally providing regional contributions of the aforementioned geophysical fields.

\section{Lagrangian drift simulation with DriftNet: multivariate extension}

DriftNet is a Convolutional Neural Network model for simulating Lagrangian drift at the sea surface from given Eulerian geophysical fields~\citep{Botvynko2023, botvynko2025neural}. Its main originality consists on considering an Eulerian representation of the Lagrangian trajectory through the Fokker-Planck equation:

\begin{equation}\label{eq:fokker-planck}
\frac{\partial}{\partial t} p_{\vec{r}}(\mathbf{x},t)=-\frac{\partial}{\partial \mathbf{x}} \left[ \mu(\mathbf{x},t)p_{\vec{r}}(\mathbf{x},t) \right]
\end{equation}

\noindent where $\mathbf{x}$ and $t$ are the spatial and temporal coordinates in the Eulerian formalism, $\vec{r}$ is the Lagrangian position, $\mu$ is the geophysical field used to advect the Lagrangian particle and $p_{\vec{r}}$ is the Eulerian representation of the Lagrangian tracer.

In practice, Driftnet considers the following latent representation for a simulated Lagrangian drift $\hat{\vec{r}}$:
\begin{equation}
\left \{
    \begin{array}{ccl}
         \mathbf{y} &=&  \mathcal{E} \left ( \mathbf{g}, \mathbf{y}_0 \right ) \\
         \hat{\vec{r}} &=&  \mathcal{M} \left (\mathbf{y} \right ) \\
    \end{array}\right.
\end{equation}
where $\mathbf{y} = \{\mathbf{y}_{t_0},\mathbf{y}_{t_0+\Delta},\ldots,\mathbf{y}_{t_0+K\Delta}\}$ is a space-time-explicit Eulerian embedding of $\vec{r}$, and $\mathbf{y}_0$ some initial Eulerian embedding to encode the initial position $\vec{r}_{t_0}$.  
%
The neural operator $\mathcal{E}$ computes $\mathbf{y}$ given geophysical conditions $\mathbf{g}$ and initial representation of the position of the particle $\mathbf{y}_0$.

\begin{figure*}[h]
    \centering
    \includegraphics[width = \textwidth]{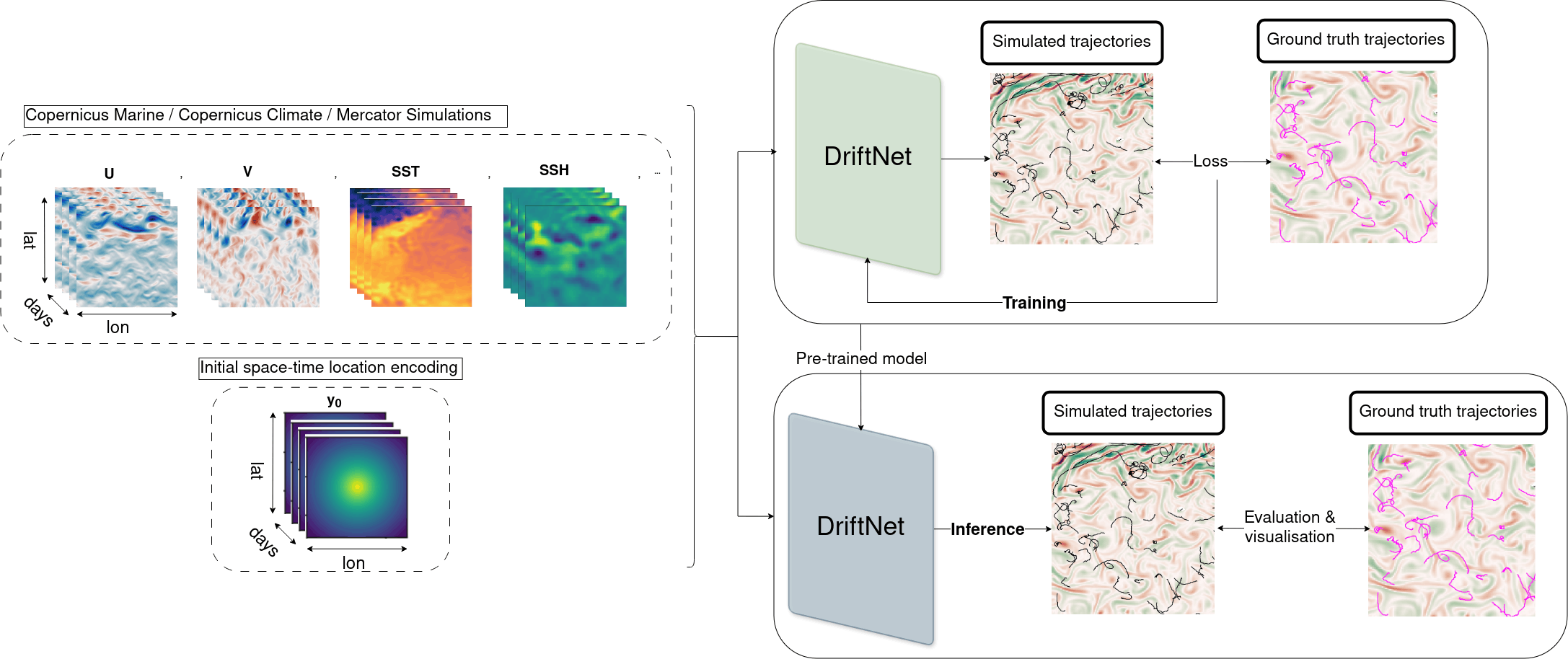}
    \caption{Training and evaluation procedures of multivariate DriftNet's extension: the input geophysical fields $\mathbf{g}$ over 9 days (here zonal $\mathbf{U}$ and meridional $\mathbf{V}$ components of the velocity field, SST and SSH) coupled to the intial spatio-temporal positional encoding $\mathbf{y_{0}}$ are fed to the DriftNet in order for it to generate the target trajectories to be compared to the ground-truth ones. Once DriftNet is trained, those geophysical fields are then used as input to it in order to simulate Lagrangian trajectories which are then evaluated on the set of test trajectories using evaluation metrics described in Section~\ref{sec:eval_metrics}.}
    \label{fig:driftnet_architecture}
\end{figure*}

Then, the operator $\mathcal{M}$ maps the latent representation $\mathbf{y}$ to the targeted Lagrangian drift $\vec{r}$. 
The analytical expression of the operator $\mathcal{M}$ for a given time step $t$ grounds on the Hadamard product between the spatially-explicit encoding $\mathbf{x}$ and $\mathbf{y}_{t}$ fields.
architecture of DriftNet is illustrated in figure~\ref{fig:driftnet_architecture}. For more details on the DriftNet architecture, modeling and learning scheme we refer the reader to~\citep{botvynko2025neural}.

DriftNet architecture can be adapted to the integration of additional geophysical input fields by simply increasing the number of channels in its first convolutional layer, see Figure \ref{fig:driftnet_architecture}. The spatially and temporally aligned fields, including SSC, SSH, SST, winds and Ekman current, are concatenated with the space-time encoding of initial particles' locations. 

\section{Experimental design}

We design the experimental setup to evaluate the performance of the proposed multivariate DriftNet model under both idealized and realistic oceanic conditions. We conduct the experiments in two dynamically distinct regions, the highly energetic Gulf Stream~\citep{dewar1989gulf} and the relatively smooth North East Pacific~\citep{pacific_dynamics}, in order to ensure a robust assessment of the model's generalizability across varying ocean regimes.

This section is organized as follows: we first describe the Eulerian datasets, including both synthetic and operational ocean and atmospheric models. Next, we present the Lagrangian datasets, covering synthetic particle trajectories and observed drifter data. We then define two benchmarks scenarios designed to evaluate the impact of geophysical variables on trajectories simulation with DriftNet under idealized and realistic conditions. Finally, we detail the evaluation metrics applied to evaluate the model's performances across the proposed benchmarks.

\subsection{Eulerian datasets}\label{sec:data_eulerian}

\textbf{Nature Run - E1 dataset}
This Eulerian dataset, called from now on E1, contains the SSC velocities, SSH and SST fields from the free run model called Nature Run. The latter was kindly provided to us by Mercator Ocean and was obtained from a high-resolution NEMO simulation with no data assimilation~\citep{le2019observation}. It is regularly-gridded with a horizontal spatial resolution of $1/12^{\circ}$ and a daily temporal resolution. This dataset involves the entire year of 2015.

\begin{table*}[h]
\caption{\textbf{Summary description of Eulerian datasets used in the current study:} \textit{Source} makes reference to the used numerical or satellite product, \textit{Variable} indicates the physical quantity of interest, \textit{Period} the studied years. The spatial and temporal resolutions of each product are also indicated. SSC - Sea Surface Currents. SSH - Sea Surface Height. SST - Sea Surface Temperature.}\label{table_dataset}
\begin{center}
\begin{tabular}{p{0.8cm}|p{5.1cm}|p{2.cm}|p{1.7cm}|p{1.6cm}|p{1.55cm}}
\hline
\textit{Name} & \textit{Source} & \textit{Variable} & \textit{Period} &  \textit{Spatial resolution} & \textit{Temporal resolution} \\
\hline
\hline
E1 & Nature Run~\citep{le2019observation} & SSC, SSH, SST & 2015 & 1/12$^{\circ}$  & 1 day\\
E2 & OSSE-based assimilated free run \citep{verrier2017assessing, verrier2018assessing} & SSC & 2015  & 1/12$^{\circ}$ & 1 day\\
E3 & GLORYS12~\citep{glo12} & SSC & 1992 - 2020 &  1/12$^{\circ}$  & 1 day \\
E3 & DUACS \citep{pujol2016duacs} & SSH & 1992 - 2020 & 1/8$^{\circ}$ & 1 day\\
E3 & OSTIA \citep{good2020current} & SST & 1992 - 2020 & 1/20$^\circ$ & 1 day \\
E3 & GLOBCURRENT \citep{rio2014beyond} & Ekman & 1992 - 2020 & 1/4$^{\circ}$ & 1 day\\
E3 & ERA5 \citep{dee2011era} & Winds & 1992 - 2020 & 1/4$^{\circ}$ & 1 day\\
\hline
\end{tabular}
\end{center}
\end{table*}

\textbf{OSSE - E2 dataset}
This dataset, now called E2, contains the SSC velocities from the Observing System Simulation Experiments (OSSEs) framework, kindly provided to us by Mercator Ocean. This experiment dynamically reconstructs the ocean state using pseudo observations sampled from E1. These pseudo observations mimic altimetry missions from Jason-3, Sentinel 3A and Sentinel 3B. Same as E1, the E2 is provided on a regular grid with 1/12° horizontal and daily temporal resolutions. This dataset covers the entire year of 2015.

\begin{figure*}[h]
    \centering
    \includegraphics[width=0.85\textwidth]{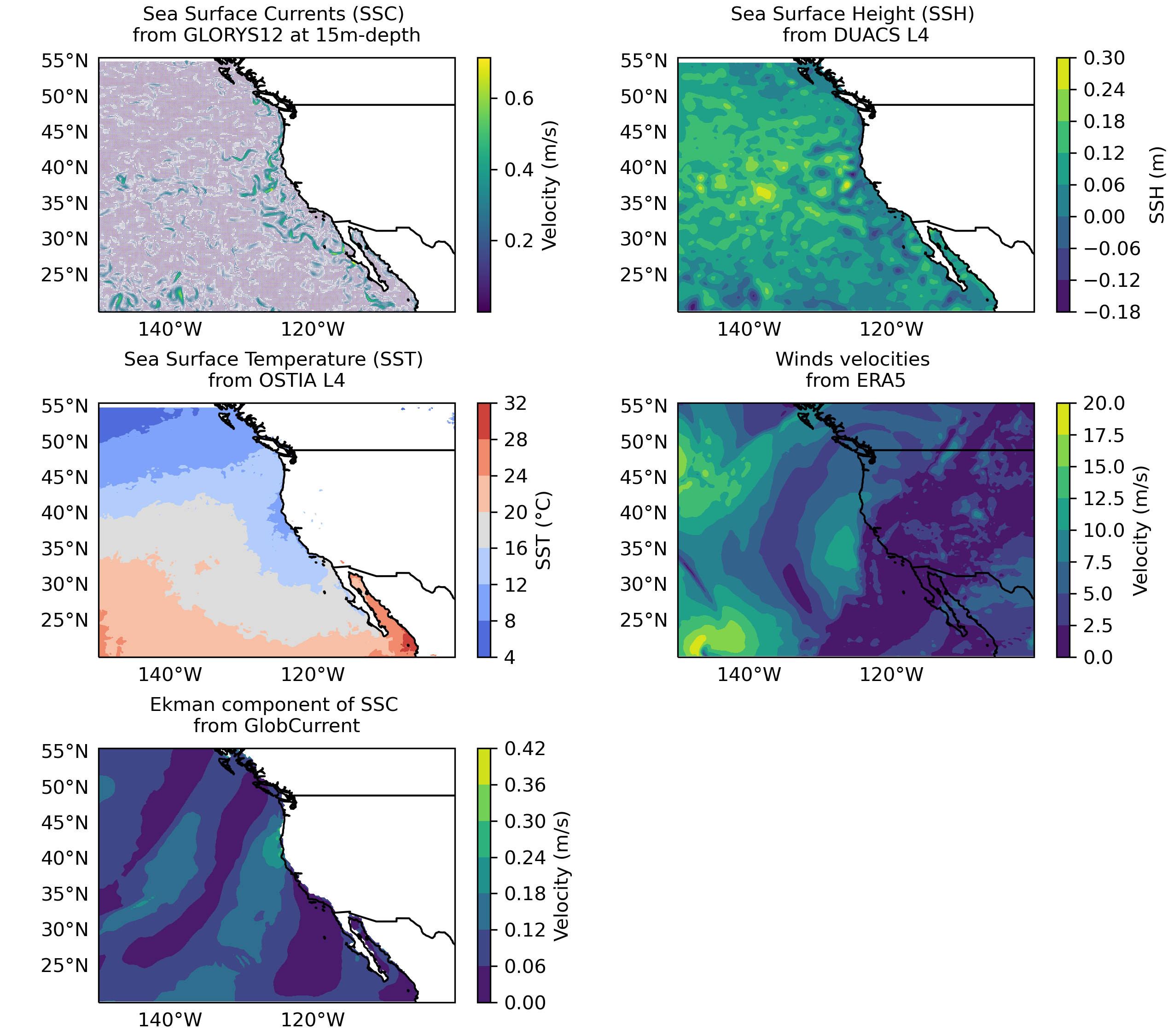}
    \caption{North East Pacific: SSC velocity in m/s, SST in $^{\circ} C$, SSH in m, wind velocity in m/s, Ekman in m/s. Snapshot on 01/06/2015 12:00:00 UTC.}
    \label{fig:all_variables_snapshot_Pacific}
\end{figure*}

\begin{figure*}[h!]
    \centering
    \includegraphics[width=0.85\textwidth]{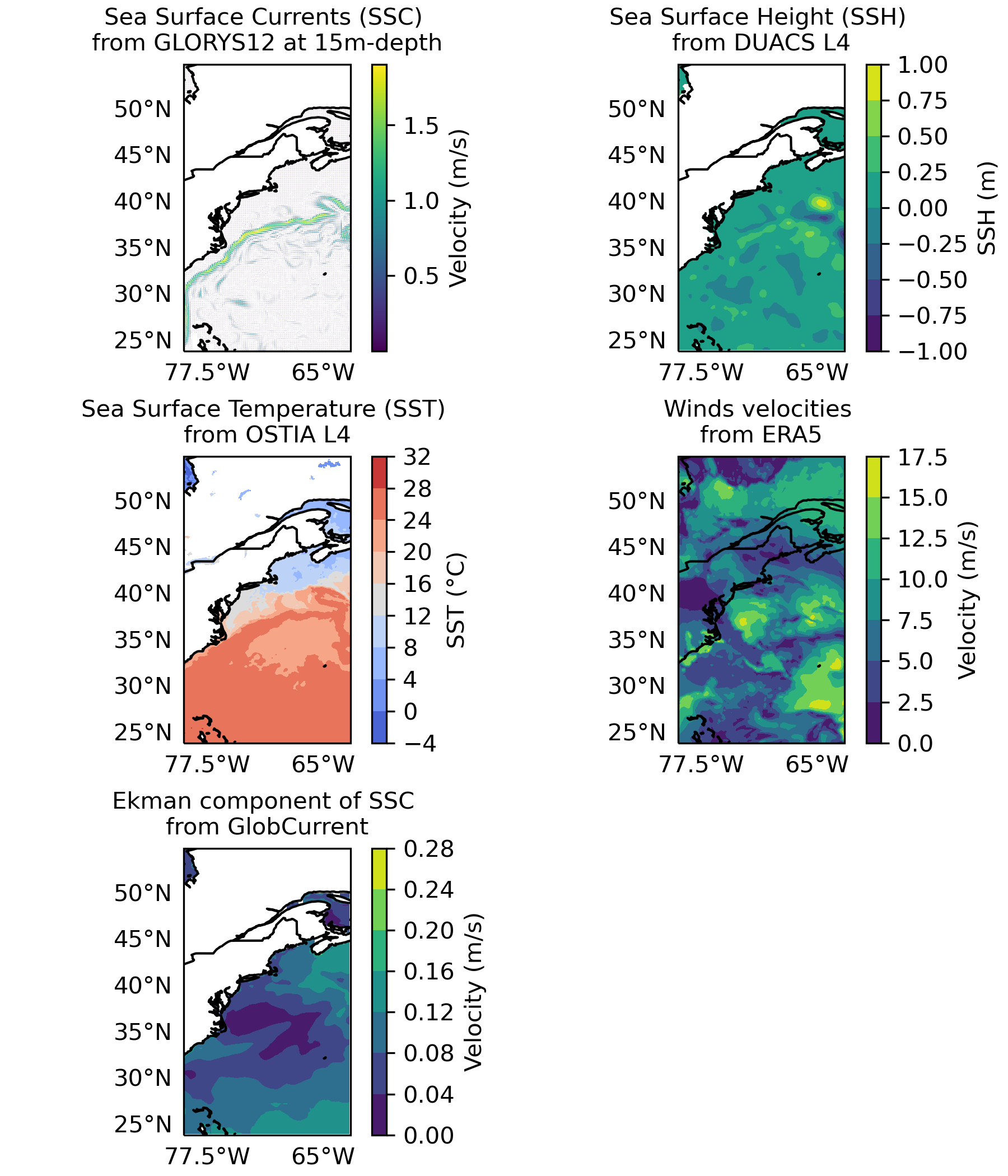}
    \caption{Gulf Stream: SSC velocity in m/s, SST in $^{\circ} C$, SSH in m, wind velocity in m/s, Ekman component in m/s. Snapshot on 01/06/2015 12:00:00 UTC.}
    \label{fig:all_variables_snapshot_GS}
\end{figure*}

\textbf{Operational ocean and atmospheric models - E3 dataset}
This real-world dataset contains the Sea Surface Currents from the global ocean physics reanalysis GLORYS12~\citep{glo12}, winds velocities from ERA5 reanalysis model \citep{dee2011era}, the Sea Surface Height from DUACS \citep{pujol2016duacs}, the Sea Surface Temperature from OSTIA \citep{good2020current} and Ekman current from GlobCurrent \citep{rio2014beyond}. To ensure data consistency, all geophysical inputs are spatio-temporally interpolated to match the resolution of GLORYS12 outputs. A snapshot for a given date of these geophysical fields can be seen in Fig.\ref{fig:all_variables_snapshot_Pacific} and Fig.\ref{fig:all_variables_snapshot_GS} for North East Pacific and Gulf Stream respectively.

The aforementioned Eulerian datasets E1 and E2 contain (2+1)d variables, where the (2+1) dimensions are two spatial and one temporal. While the SSH and SST are monovariate variables, the SSC is a multivariate variable with zonal and meridional composants. The E3 dataset contains a set of (2+1)d arrays of SSC, SSH, SST, winds velocities and Ekman. All datasets are subject of pre-processing procedure consisting in spatio-temporal alignment and filling the missing values with zeros. 

\subsection{Lagrangian datasets}\label{sec:data_lagrangian}

In this study, we refer to four Lagrangian datasets, see Table \ref{table_dataset_lagrangian}, coming from Lagrangian simulations and real surface drifters dataset. Datasets L1 and L2 correspond to Lagrangian trajectories simulated with Ocean Parcels. We configure Ocean Parcels to use Runge-Kutta 4 integration scheme with integration step of 5 minutes, applying linear interpolation scheme and no diffusive component. The duration of each trajectory simulation is defined as 9 days, and the output trajectories are provided with the regular temporal sampling of one position per 6 hours, consequently containing a total amount of 36 time steps. Dataset L3 contains real surface drifters trajectories of 9 days each with a homogeneous time sampling of 6h. 

\textbf{Nature Run-based: L1 dataset}
This dataset is composed of synthetic particles advected using the SSC from dataset E1. For each case-study region, 15 particles were randomly seeded in space every 7 days during the period of 2015, amounting to a total of 7,900 particles per region. This sampling strategy was empirically chosen in order to prepare a consistent database containing an amount of particles big enough for the data-driven method to achieve a relevant level of performance, while entirely covering the spatio-temporal range of the study. These trajectories will be used as ground-truth in the Benchmark B1 described further in Section \ref{sec:benchmarks}.

\textbf{OSSE-based: L2 dataset}
This dataset contains synthetic particles advected using the SSC from dataset E2. It contains 7,900 particles deployed at the exact same locations and times as in the dataset L1 described above. 

\textbf{Drifters database from CMEMS: L3 dataset}
This dataset contains trajectories from CMEMS drifters~\citep{etienneglobal} in the North East Pacific and in the Gulf Stream regions, from 1992 to 2020. The corresponding trajectories are divided into 9-days non-overlapping segments, each segment starting at 12:00:00 UTC. Drifters are filtered in order to keep only those which did not lose their drogue during their path. 

\begin{table}
\caption{\textbf{Summary description of Lagrangian datasets used in the current study:} \textit{Source} makes reference to the used numerical or satellite product, \textit{Variable} indicates the physical quantity of interest, \textit{Period} the studied years. The temporal resolution of each product is also indicated.}\label{table_dataset_lagrangian}
\begin{center}
\begin{tabular}{p{1.2cm}|p{5.5cm}|p{1.6cm}|p{1.7cm}|p{2.75cm}}
\hline
\textit{Name} & \textit{Source} & \textit{Variable} & \textit{Period}  & \textit{Temporal } \\
\textit{} & \textit{} & \textit{} & \textit{}  & \textit{resolution } \\
\hline
\hline
L1 & Ocean Parcels simulation using E1 SSC & Position  & 2015  &  6 hours \\
L2 & Ocean Parcels simulation using E2 SSC & Position &  2015 &  6 hours \\
L3 & CMEMS drifters~\citep{etienneglobal} & Position &  1992 - 2020 &  6 hours \\
\hline
\end{tabular}
\end{center}
\end{table}

\subsection{Benchmarks} \label{sec:benchmarks}

In this study we asses the impact of different geophysical variables using DrifNet to simulate Lagrangian drift. We use a combination of synthetic and observed datasets to construct two benchmark scenarios, and we conduct all experiments in two case-study regions: North East Pacific and Gulf Stream. In the first, Benchmark B1, we assess DriftNet using fully simulated data and we systematically vary the input geophysical fields of DriftNet: starting from a baseline using only Sea Surface Currents and then incorporating Sea Surface Height, Sea Surface Temperature, and their combinations in order to quantify their influence on trajectory simulation accuracy. Then, Benchmark B2 extends the previous experiment to real-world conditions. 

\textbf{Benchmark B1} serves as an idealized case designed to evaluate the impact of numerical geophysical fields on the accuracy of the trajectories simulated with DriftNet. In this benchmark, we use numerically simulated trajectories from L1 dataset as reference, and Eulerian geophysical fields from the Nature Run (E1) and OSSE (E2) datasets as input to DriftNet. 

We consider the following inputs combinations in order to conduct the training and metrics assessment: 
\begin{itemize}
    \item \textbf{SSC from E2}: baseline Driftnet configuration
    \item \textbf{SSH from E1}: in order to asses solely the impact of 'fully observed' SSH from Nature Run.
    \item \textbf{SST from E1}: in order to asses solely the impact of 'fully observed' SST from Nature Run. 
    \item \textbf{SSC from E2 combined to SSH from E1}: in order to assess the impact of synergistic combination of 'fully observed' SSH from Nature Run and assimilated SSC from OSSE. 
    \item \textbf{SSC from E2 combined to SST from E1}: in order to assess the impact of synergistic combination of 'fully observed' SST from Nature Run and assimilated SSC from OSSE. 
    \item \textbf{SSC from E2 combined to SSH and SST from E1}: in order to assess the impact of all previously mentioned variables combined together.
\end{itemize}

A total amount of 7900 time series of the aforementioned Eulerian fields E1 and E2 and the corresponding trajectories L1 and L2 are randomly split into training (80\%), validation (10\%) and test (10\%) subsets.

\textbf{Benchmark B2} extends the analysis from previous benchmark to a real-world setting. We train the mutlivariate extension of DriftNet to reproduce observed trajectories from the CMEMS drifter database (L3), using geophysical fields from operational ocean and atmospheric models (E3). These fields include SSC from GLORYS12, SSH from DUACS \citep{taburet2019duacs}, SST from OSTIA \citep{good2020current}, wind velocities from ERA5 \citep{dee2011era}, and Ekman currents from GLOBCURRENT \citep{cancet2019evaluation}.

We consider the following inputs combinations in order to conduct the training and metrics assessment: 
\begin{itemize}
    \item \textbf{SSC from GLORYS12}: baseline Driftnet configuration.
    \item \textbf{SSH from DUACS}: in order to asses solely the impact of gap-free altimetry observations.
    \item \textbf{SSC from GLORYS12 combined to SSH from DUACS}: in order to assess the impact of the combination of gap-free altimetry observations and ocean currents from operational assimilated reanalysis model. 
    \item \textbf{SSC from GLORYS12 combined to SST from OSTIA}: in order to assess the impact of the combination of gap-free temperature observations and ocean currents from operational assimilated reanalysis model. 
    \item \textbf{SSC from GLORYS12 combined to winds velocities from ERA5}: in order to assess the impact of the combination of winds velocities from operational assimilated reanalysis atmospheric model and ocean currents from operational assimilated reanalysis model. 
    \item \textbf{SSC from GLORYS12 combined to winds velocities from ERA5 and SSH from DUACS}: in order to assess the impact of the combination of winds velocities from operational assimilated reanalysis atmospheric model, ocean currents from operational assimilated reanalysis model and gap-free altimetry observations.
    \item \textbf{SSC from GLORYS12 combined to Ekman current from GLOBCURRENT}: in order to assess the impact of modelled Ekman currents and ocean currents from operational assimilated ocean reanalysis model.
\end{itemize}

A total amount of 12.850 time series of the aforementioned Eulerian datasets E3 and the corresponding trajectories L3 are randomly split into training (80\%), validation (10\%) and test (10\%) subsets.

\subsection{Evaluation Metrics}\label{sec:eval_metrics}
In this study, the performance is evaluated using:

\begin{itemize}

\item the mean Euclidean distance at the last time step (9-th day) in kilometers, that we denote $D$:

\begin{equation}\label{eq:mean_distance_eucl}
    D = \frac{1} { N_{T} } \cdot \sum_{i = 0}^{N_{T}}d_{i_{9}}
\end{equation}

\noindent with $N_T$ the number of trajectories and $d_{i_{9}}$ defined as: \begin{equation}\label{eq:distance_eucl}
    d_{i_{j}} = \sqrt{  (\vec{r}_{R}(\vec{r}_{0,i},j\Delta) - \vec{r}_{S}(\vec{r}_{0,i},j\Delta))^2}
\end{equation}
\noindent where $\vec{r}_{S}$ is simulated trajectory, $\vec{r}_{R}$ the reference one and $j=9$.

\item mean Normalized Cumulative Lagrangian Separation \citep{liu2011evaluation}, or mean Liu index $\mathcal{L}_{Liu}$, without units:

\begin{equation}\label{eq:lossliu}
    \mathcal{L}_{Liu}=\frac{1}{N_{T}} \sum_{i=1}^{N_{T}} \frac{\sum_{j=1}^{K} d_{i_{j}}}{\sum_{j=1}^{K} l_{i_{j}}}
\end{equation}

\noindent where $l_{i_{j}}$ is the cumulative length of reference trajectory $i$ between its positions at time steps $j$ and $j-1$.

\item the mean error between autocorrelation functions in meridional and zonal components of the trajectories' velocities, $R_{\vec{u}}(\tau)$ and $R_{\vec{v}}(\tau)$ respectively ~\citep{lagr_time_scale, kang2005scale, wunsch1999interpretation}:

\begin{equation}
	\Delta R_{\vec{u}} = \langle \vert \overline R_{\vec{u}_{R}}(\tau) - \overline R_{\vec{u}_{S}}(\tau) \vert \rangle
	 \label{eq:mean_lagr_time_scale} 
\end{equation}
\noindent 
\noindent where $\langle \rangle$ is the average across  temporal scale dimension ($\tau$), $\overline R_{\vec{u}_{R}}(\tau)$ and $\overline R_{\vec{u}_{S}}(\tau)$ are respectively the ensemble mean of the reference and simulated normalized autocorrelation functions of the Lagrangian velocity, defined as:  $R_{\vec{u}}(\tau)=\frac{1}{9\cdot \Delta} \cdot \sum_{t=0}^{9} \vec{u}(t\Delta+\tau) \vec{u}(t\Delta)$. The same definition is used for the meridional velocity.

\end{itemize}

The mean euclidean distance and the mean Liu index (NCLS) are computed between the reference and simulated trajectories and averaged over the whole ensembles of the test datasets. The autocorrelation functions' errors are estimated as the differences between the average autocorrelation functions over the ensembles of trajectories from test datasets. 

\section{Results}
In this section, we evaluate the impact of the different geophysical variables on the quality of the DriftNet Lagrangian simulations in idealized, benchmark B1, and real-world, benchmark B2, settings.

\subsection{Results for Benchmark B1}

Table~\ref{t5} shows the performance of DriftNet to generate Lagrangian trajectories with different combinations of geophysical input fields in a fully numerical benchmark (B1) and in both, North East Pacific and Gulf Stream, regions. 

The outperformance of Driftnet compared to Ocean Parcels simiulations has been demonstrated in \citep{botvynko2025neural}. Nevertheless, it is important to mention that L2 and L1 trajectories are generated by the same advection-based algorihtm, thus producing trajectories with similar roughness, which explains the better performance of Ocean Parcels wrt DriftNet across the autocorrelation metric. However, this roughness is not representative of the real-world trajectories while comparing Ocean Parcels simulations as defined in Section \ref{sec:data_lagrangian} to surface drifters \citep{botvynko2025neural}. As consequence, we do not further analyze Ocean Parcels performance in the following. For further details on the comparison between Ocean Parcels simulations and surface drifters we refer the reader to \citep{botvynko2025neural}.

In the North East Pacific the combination of fully observed SSH and assimilated SSC provides the best Lagrangian trajectories and allows to reduce $D$ by over 55\%, $\mathcal{L}_{Liu}$ by over 60\%, $\Delta R_{\vec{u}}$ and $\Delta R_{\vec{v}}$ by over 60\% and 50\% respectively compared to the baseline DriftNet configuration, see Table \ref{t5}. When providing only SSH input to the DriftNet, the separation distance $D$ is reduced by more than 31\%, the mean $\mathcal{L}_{Liu}$ is reduced by more than 54\%, $\Delta R_{\vec{u}}$ and $\Delta R_{\vec{v}}$ are both reduced by 50\% and 42\% respectively. These results illustrate the high amount of information on surface dynamics contained in the SSH. However, in the case of SST input only, the mean separation distance is increased by 9\%, $\Delta R_{\vec{u}}$ is increased by 16\%, while the mean $\mathcal{L}_{Liu}$ is reduced by almost 3\% and $\Delta R_{\vec{v}}$ remains at the same level as the baseline. Furthermore, combining SST and SSC increases the mean separation distance $D$ by almost 4\% and $\Delta R_{\vec{u}}$ and $\Delta R_{\vec{v}}$ errors are increased by 25\% and 33\% respectively, while the mean $\mathcal{L}_{Liu}$ is reduced by more than 4\%. The combination of all previously mentioned variables, SSC, SSH and SST, leads to the reduction of mean separation distance $D$ by almost 52\%, the reduction of mean $\mathcal{L}_{Liu}$ by more than 57\%, and the errors of  $\Delta R_{\vec{u}}$ and $\Delta R_{\vec{v}}$ are both reduced by almost 42\% and 33\% respectively.
Similarly in the Gulf Stream, the best performances across the Lagrangian metrics are obtained when combining SSH from a free run and SSC from its assimilated version. In this configuration we report a relative gain of almost 18\% for mean separation distance $D$, more than 17\% for mean $\mathcal{L}_{Liu}$, while $\Delta R_{\vec{u}}$ error is reduced by 50\% and $\Delta R_{\vec{v}}$ error is reduced by more than 57\% compared to the baseline configuration, see Table \ref{t5}. When the SSH is provided as input of DriftNet alone, the mean $D$ is increased by 4\%, the mean $\mathcal{L}_{Liu}$ is increased by 11\%, while $\Delta R_{\vec{u}}$ and $\Delta R_{\vec{v}}$ are reduced by more than 56\% and 57\% respectively. While providing SST only as input of DriftNet, the mean separation distance is increased by almost 10\%, the mean $\mathcal{L}_{Liu}$ is increased by 22\%, while $\Delta R_{\vec{u}}$ and $\Delta R_{\vec{v}}$ are reduced by 50\% and more than 42\% respectively. While combining SST to SSC, the mean separation distance $D$ is increased by 9\%, the mean $\mathcal{L}_{Liu}$ is increased by 25\% and the $\Delta R_{\vec{u}}$ and $\Delta R_{\vec{v}}$ errors are reduced by more than 37\% and 21\% respectively. Combining all the variable together, SSC, SSH and SST, increases the mean separation distance $D$ by 10\%, the mean $\mathcal{L}_{Liu}$ by almost 16\%, and reduces the $\Delta R_{\vec{u}}$ and $\Delta R_{\vec{v}}$ errors by almost 19\% and 14\% respectively. 

\begin{figure}
\centering
\begin{subfigure}{0.75\textwidth}
\includegraphics[width=\textwidth]{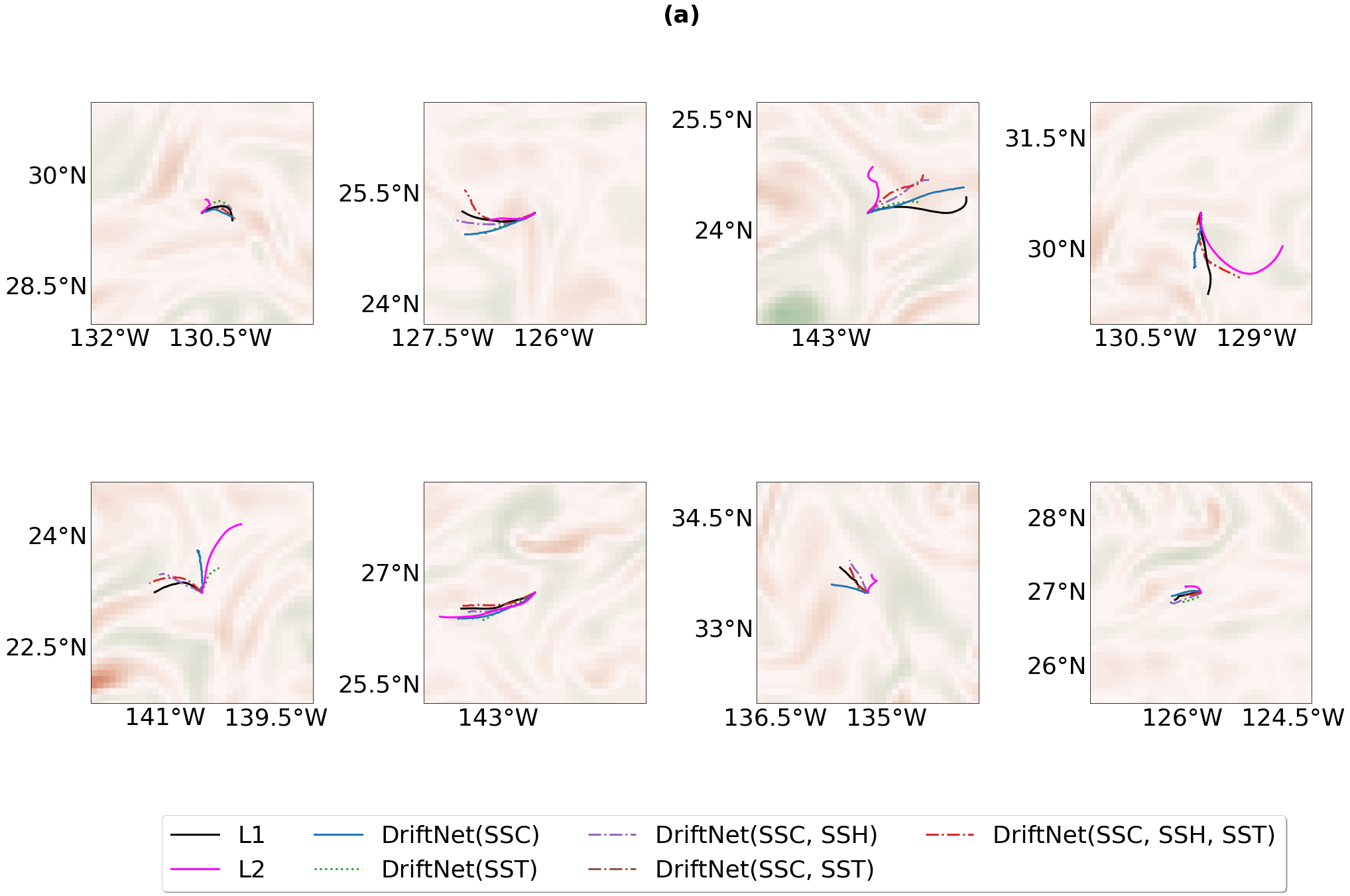}
\end{subfigure}
\begin{subfigure}{0.75\textwidth}
\vspace{0.6cm}
\includegraphics[width=\textwidth]{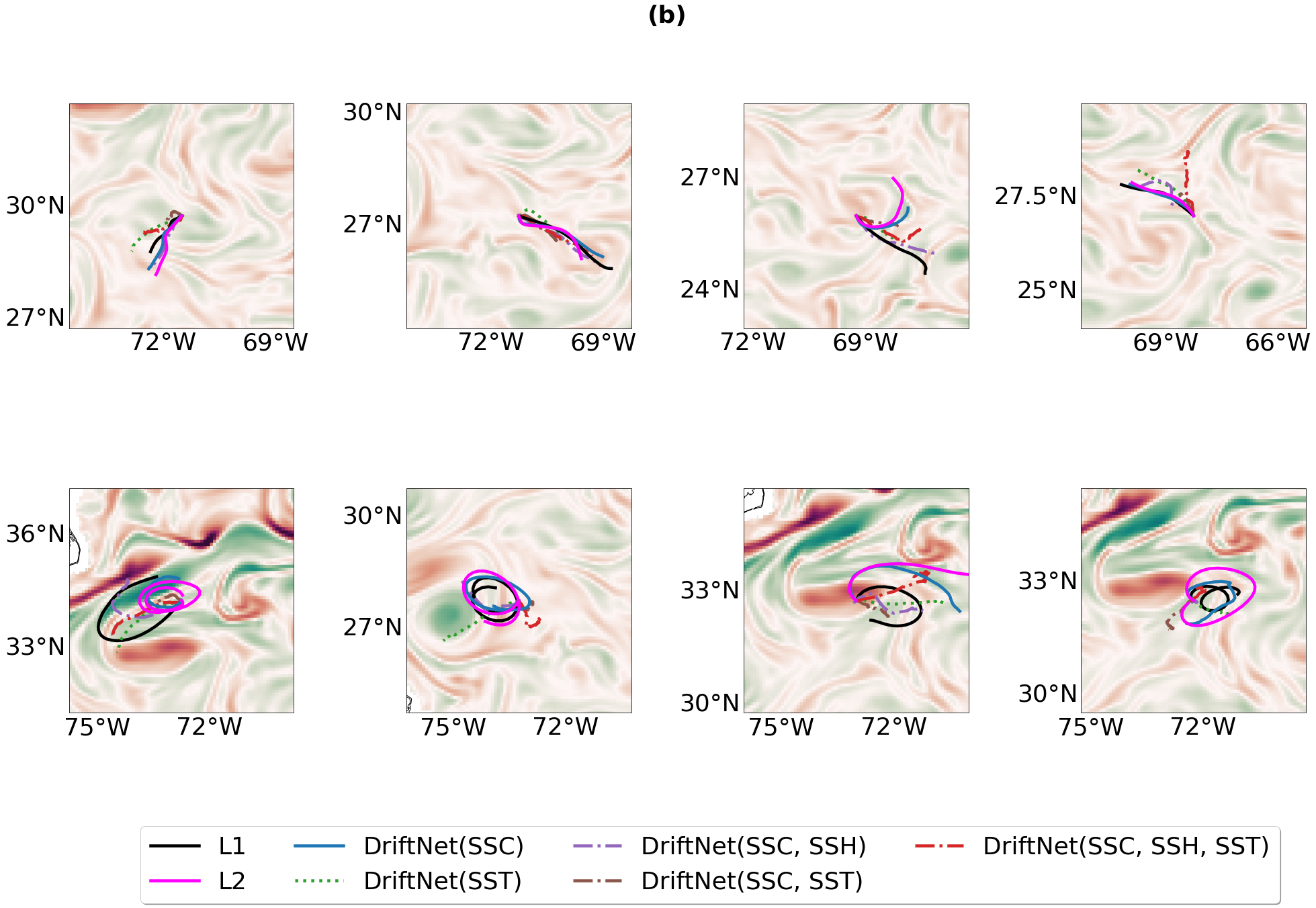}
\end{subfigure}
\caption{\textbf{Examples of simulated trajectories for Benchmark B1}: Panel \textbf{(a)}: North East Pacific, Panel \textbf{(b)}: Gulf Stream. The reference trajectories simulated with Ocean Parcels using Nature Run SSC L1, the baseline ones L2 simulated using SSC from OSSE. Eight randomly-selected trajectories are superimposed to the mean relative vorticity of the Nature Run SSC.}\label{fig:trajectories_B1}
\end{figure}

In both regions, we observe that DriftNet provided with SSC and SSH input matches the best to the reference trajectories L1, see Fig.\ref{fig:trajectories_B1}. These results are in agreement with the ability of SSH products to improve SSC representation~\citep{ballarotta2022improved, liu2014evaluation}. By contrast, SST does not provide comparable gains. The degradation of performance associated with the combination of SSH, SST and SSC in the Gulf Stream deserves a deeper investigation. This phenomenon is probably due to the misalignement of mesoscale features in SST and SSC fields \citep{seo2016eddy, aguedjou2023imprint, tang2022mesoscale}, or to the overfitting problem, suggesting the need for a different training strategy. 

\begin{table}[t]
\caption{\textbf{Performance of DriftNet with different Eulerian geophysical input fields and Ocean Parcels (L2 dataset) in Benchmark B1 with respect to the reference trajectories (L1 dataset).} We benchmark DriftNet scheme using as inputs data-assimilation-based SSC from dataset E2 and idealized 'fully observed' SSH and SST fields from dataset E1. We illustrate performance in two study regions: North East Pacific and Gulf Stream. The evaluation metrics are the last time step mean separation distance ($D$), mean Liu index ($\mathcal{L}_{Liu}$) and error between autocorrelation functions of zonal and meridional components respectively ($\Delta R_{\vec{u}}$ and $\Delta R_{\vec{v}}$). Bold indicates the best performance.}\label{t5}
\begin{center}
\begin{tabular}{ccccc|cccc}
\hline
&  \multicolumn{3}{c}{\textbf{North East Pacific}}  & & \multicolumn{4}{c}{\textbf{Gulf Stream}}   \\
 & $D$, km & $\mathcal{L}_{Liu}$ & $\Delta R_{\vec{u}}$ & $\Delta R_{\vec{v}}$ &  $D$, km & $\mathcal{L}_{Liu}$ & $\Delta R_{\vec{u}}$ & $\Delta R_{\vec{v}}$ \\
\hline
L2 & 53.8 & 0.87 &  \textbf{0.0086} & \textbf{0.016} & 168.3 & 0.75 &  0.01  & 0.02  \\
\hline
SSC & 45.4 & 0.70  & 0.12  & 0.12 & 154.6 & 0.63 &  0.16 & 0.14 \\
SSH & 31.2 & 0.32 & 0.06 & 0.07 & 160.3 & 0.70 & \textbf{0.07} & \textbf{0.06} \\
SST & 49.6 & 0.68 & 0.14 & 0.12 & 170.3 & 0.77 & 0.08 & 0.08 \\
SSC, SSH & 22.2 & \textbf{0.29} & \textbf{0.05} & \textbf{0.06} & \textbf{127.0}  & \textbf{0.52} & 0.08   & \textbf{0.06}\\
SSC, SST & 47.1 & 0.67 & 0.15 & 0.16 & 168.9  & 0.79 & 0.10 & 0.11\\
SSC, SSH, SST & \textbf{22.0} & 0.30 & 0.07 & 0.08 & 170.3 & 0.73 & 0.13 & 0.12 \\
 \hline
\end{tabular}
\end{center}
\end{table}

\subsection{Results for Benchmark B2} 

Table~\ref{t6} shows the performance of DriftNet to generate Lagrangian trajectories with different combinations of geophysical input fields in a real-world benchmark (B2) and in both, North East Pacific and Gulf Stream, regions. 

In the North East Pacific the combination of GLORYS12 SSC and satellite-derived SSH reduces the mean separation distance by almost 4\%, the mean $\mathcal{L}_{Liu}$ by more than 5\%, and the $\Delta R_{\vec{u}}$ error by 40\%, while the $\Delta R_{\vec{v}}$ error remains at the same level, all compared to the baseline. When providing GLORYS12 SSC and SST from OSTIA product, the mean $D$ is slightly increased by almost 2\%, the mean $\mathcal{L}_{Liu}$ remains at the same level and the errors in autocorrelations functions are reduced by 40\% and 33\% for zonal and meridional components respectively. The combination of winds velocities from ERA5 and GLORYS12 SSC increases the mean $D$ by 1\%, and keeps the mean $\mathcal{L}_{Liu}$ at the same level as the baseline configuration. The SSC-Ekman combination increases the mean $D$ by 3\%, and reduces the autocorrelation error in the zonal component by 40\%. The inclusion of wind velocities from ERA5 does not significantly affect performance, while the addition of Ekman current results in a modest increase in $D$ and a 40\% reduction in $\Delta R_{\vec{u}}$. The best performance in terms of $D$ is achieved with the SSC-SSH-wind combination, which decreases $D$ by almost 4\% and moderately reduces $\mathcal{L}_{Liu}$.

\begin{table}
\caption{\textbf{Performance of DriftNet with different Eulerian geophysical input fields from remote sensing and reanalysis models with respect to the reference trajectories (L3 dataset) in Benchmark B2.} We indicate the considered input data: SSC, namely the velocity fields U and V from the GLORYS12 reanalysis~\citep{lellouche_glorys12}, complemented by satellite-derived optimally-interpolated SSH fields~\citep{pujol2016duacs}, SST fields \citep{good2020current}, winds fields \citep{dee2011era} and Ekman \citep{rio2014beyond}. We report the performance metrics in the two study regions: North East Pacific and Gulf Stream. The evaluation metrics are the last time step mean separation distance ($D$), mean Liu index ($\mathcal{L}_{Liu}$) and mean absolute error between autocorrelation functions ($\Delta R_{\vec{u}}$ and $\Delta R_{\vec{v}}$). Bold indicates the best performance.}\label{t6}
\centering
\begin{tabular}{ccccc|cccc}
\hline
&  \multicolumn{3}{c}{\textbf{North East Pacific}}  & & \multicolumn{4}{c}{\textbf{Gulf Stream}}   \\
\hline
 & $D$, km & $\mathcal{L}_{Liu}$ & $\Delta R_{\vec{u}}$ & $\Delta R_{\vec{v}}$ &  $D$, km & $\mathcal{L}_{Liu}$ & $\Delta R_{\vec{u}}$ & $\Delta R_{\vec{v}}$ \\
\hline
SSC & 62.4 & 0.55 & 0.05 & 0.03 &  118.4 & 0.6 &  0.08  & 0.08 \\
SSC, SSH & 60.2 & 0.52 & \textbf{0.03}  & 0.03 & 122.4 & 0.61 & 0.03 & \textbf{0.03} \\
SSC, SST & 63.6 & 0.55 &  \textbf{0.03} & \textbf{0.02} & \textbf{103.2} & \textbf{0.51} & 0.04 & \textbf{0.03} \\
SSC, Wind & 63.2 & 0.55 &  0.06 & 0.06 & 136.4 & 0.7 & 0.08 & 0.08\\
SSC, SSH, Winds &  \textbf{60.1} &  0.52 & 0.05   & 0.04 & 136.2 & 0.65 & 0.1 & 0.1 \\
SSC, Ekman & 64.3  & 0.55 & 0.06 & 0.03 & 117.1 & 0.58 & 0.03 &  0.04 \\
SSC, SSH, Ekman & 62.1  & \textbf{0.51} & \textbf{0.03} & 0.04 & 110.9 & 0.54 & 0.05 & 0.05  \\

\hline
\end{tabular}
\end{table}

In the Gulf Stream, the best performance compared to the baseline are obtained with the SST-SSC combination, where the mean separation distance is reduced by 15\%, the Liu index is reduced by 15\% and the autocorrelation error of the meridional velocity by 62\%. On the other hand, Ekman also reduces of the mean separation distance by more than 6\%, the Liu index by 10\% and the autocorrelation errors by more than 60\% in both components compared to the baseline input configuration. In contrast, combining winds with SSC either alone or in combination with SSH gives inferior performance, with increases in separation distance and negligible improvements in the Liu index. These results suggest that wind-driven processes may contribute less to surface trajectory dynamics in the Gulf Stream than geostrophic currents, or may introduce stochastic variability that is not well captured in deterministic simulations.

\begin{figure}
     \centering
     \begin{subfigure}{0.85\textwidth}
         \includegraphics[width=\textwidth]{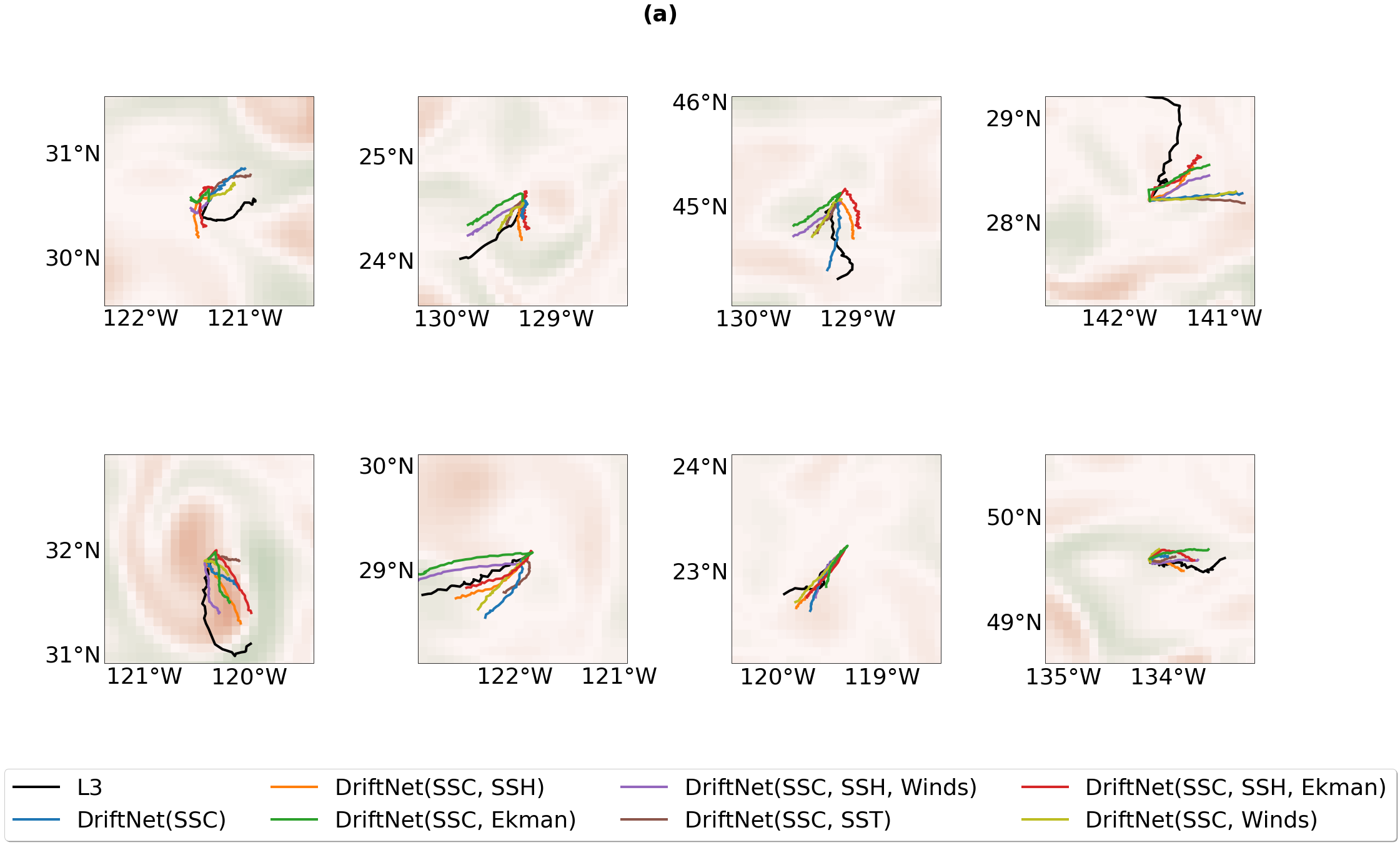}
     \end{subfigure}
     \vspace{0.6cm}
     \begin{subfigure}{0.85\textwidth}
         \includegraphics[width=\textwidth]{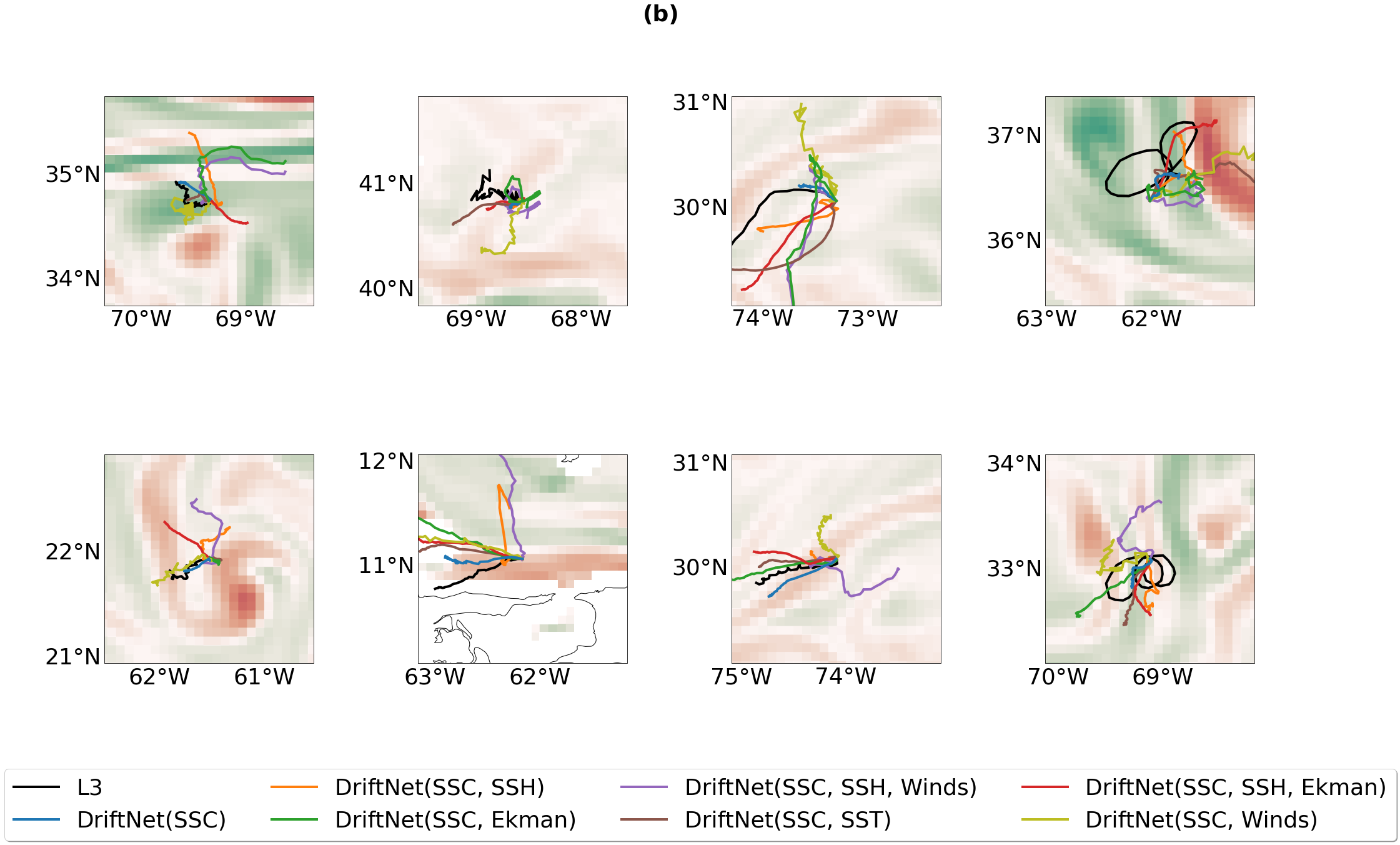}
     \end{subfigure}
     \hfill
        \caption{\textbf{Examples of simulated trajectories for Benchmark B2}: Panel \textbf{(a)}: North East Pacific, Panel \textbf{(b)}: Gulf Stream. We depict real drifters trajectories in black (L3) and trajectories simulated with DriftNet using various input geophysical fields.  Eight randomly-selected trajectories are superimposed to the mean relative vorticity of GLORYS12.}
        \label{fig:trajectories_B2}
\end{figure}

In the North East Pacific there is a clear beneficial impact of SSC-SSH combination, giving trajectories ending up the closest to the drifters' ones at the last time step in most cases, see Fig.\ref{fig:trajectories_B2}.a. On the other hand, in the Gulf Stream, we observe the beneficial impact of SSC-SST combination, see Fig.\ref{fig:trajectories_B2}.b. Interestingly, including SSH in the Gulf Stream does not consistently improve simulations, while SST seems to be more informative to DriftNet applied in this region.

\section{Conclusion and Discussion}
This study presents a multivariate version of DriftNet~\citep{botvynko2025neural} able to deal with multiple Eulerian geophysical fields (Sea Surface Height, Sea Surface Temperature, wind velocity, and Ekman current) as input. We study how different combinations of these fields affect the accuracy of DriftNet Lagrangian trajectory simulations. We focus on two case studies: a fully numerical benchmark B1 and a real-world experiment B2. In both benchmarks, we evaluated the performance of different input combinations in two contrasting oceanic regions: the North East Pacific and the Gulf Stream.

In the fully numerical benchmark B1, the combination of fully observed SSH and assimilated SSC consistently produced the most accurate simulations in both regions. This combination significantly reduced the mean separation distance and the NCLS, and improved autocorrelation-based metrics. The strong performance of SSH in this context highlights its role in informing large-scale geostrophic dynamics that are otherwise underrepresented in assimilated SSC products \citep{chelton2011global, dufau2016mesoscale}. These findings are consistent with theoretical expectations and recent studies, suggesting that SSH fields help neural models to better reconstruct mesoscale structures \citep{martin2023synthesizing, le2025vardyn}.

In the real-world experiment B2, there is no clear optimal configuration. The best combination of fields should be chosen depending on specific applications requirements and region. In the North East Pacific, the lowest mean separation distance and NCLS are obtained when combining DUACS SSH and GLORYS12 SSC with ERA5 wind fields or GlobCurrent-derived Ekman currents \citep{taburet2019duacs, glo12, dee2011era, cancet2019evaluation}. This combination puts in evidence the strong interplay of large-scale geostrophic flow and wind-driven surface dynamics in shaping Lagrangian motion in this region \citep{niiler1995wind, onink2019role}. It is not clear however which combination of variables impacts the most the autocorrelation metrics thus suggesting the need for further investigation. In contrast, in the Gulf Stream, characterized mostly by intense mesoscale variability, the SSC-SST combination appears to be the most informative to the DriftNet. This suggests that SST plays an important role in shaping surface currents in this highly dynamic region which are potentially underestimated in existing SSC products \citep{gula2015gulf, martin2023synthesizing}. Interestingly, in the Gulf Stream, including SSH did not consistently improve trajectory simulations. Eventually, this could reflect the highly energetic and variable dynamics of the Gulf Stream, where rapid mesoscale features produce SSH patterns that may be spatially or temporally misaligned with surface currents, limiting their contribution to Lagrangian drift simulation \citep{chelton2011global, tang2022mesoscale}. Finally, in both regions the combination of ERA5 wind fields and SSC degraded performance, possibly due to unresolved stochastic variability \citep{lumpkin2010surface}.

In further studies, the accuracy of DriftNet Lagrangian simulations should be evaluated by incorporating: (i) other geophysical variables such as ocean color or waves \citep{liu2025detection, poulain2009wind, ozgokmen2000predictability}; (ii) improved altimetry products such as those obtained from advanced neural mapping schemes \citep{morrow2012recent, ballarotta2019resolutions, ballarotta2022improved, fablet2023multimodal}, or from recent wide-swath sensors \citep{fu2014transition, le2021joint}. Additionally, DriftNet could also benefit from ensemble modelling, by either noising the inputs or extending its neural architecture to a generative one.

\section*{Acknowledgments}

We gratefully acknowledge the Mercator Ocean International and Copernicus Marine Environment Monitoring Service (CMEMS) for providing access to the comprehensive datasets used in this study, which was instrumental in our research efforts. Special thanks to Mounir Benkiran for generously providing Nature Run and OSSE data, which significantly enriched our analysis and insights. This work was supported by LEFE program (LEFE MANU and IMAGO projects IA-OAC), CNES (OSTST DUACS-HR and SWOT ST DIEGO) and ANR Projects Melody (ANR-19-CE46-0011) and OceaniX (ANR-19-CHIA-0016). It benefited from HPC and GPU resources from Azure (Microsoft Azure grant) and from GENCI-IDRIS (Grant 2021-101030).

\bibliography{bibliography}

\begin{thebibliography}{}

\bibitem[Aguedjou et~al., 2023]{aguedjou2023imprint}
Aguedjou, H. M.~A., Chaigneau, A., Dadou, I., Morel, Y., Balo{\"\i}tcha, E.,
  and Da-Allada, C.~Y. (2023).
\newblock Imprint of mesoscale eddies on air-sea interaction in the tropical
  atlantic ocean.
\newblock {\em Remote Sensing}, 15(12):3087.

\bibitem[Ballarotta et~al., 2019]{ballarotta2019resolutions}
Ballarotta, M., Ubelmann, C., Pujol, M.-I., Taburet, G., Fournier, F., Legeais,
  J.-F., Faug{\`e}re, Y., Delepoulle, A., Chelton, D., Dibarboure, G., et~al.
  (2019).
\newblock On the resolutions of ocean altimetry maps.
\newblock {\em Ocean science}, 15(4):1091--1109.

\bibitem[Ballarotta et~al., 2022]{ballarotta2022improved}
Ballarotta, M., Ubelmann, C., Veillard, P., Prandi, P., Etienne, H., Mulet, S.,
  Faug{\`e}re, Y., Dibarboure, G., Morrow, R., and Picot, N. (2022).
\newblock Improved global sea surface height and currents maps from remote
  sensing and in situ observations.
\newblock {\em Earth System Science Data Discussions}, 2022:1--32.

\bibitem[Botvynko et~al., 2023]{Botvynko2023}
Botvynko, D., Granero-Belinchon, C., van Gennip, S., Benzinou, A., and Fablet,
  R. (2023).
\newblock Deep learning for lagrangian drift simulation at the sea surface.
\newblock In {\em {ICASSP} 2023 - 2023 {IEEE} International Conference on
  Acoustics, Speech and Signal Processing ({ICASSP})}, pages 1--5.

\bibitem[Botvynko et~al., 2025]{botvynko2025neural}
Botvynko, D., Granero-Belinchon, C., Van~Gennip, S., Benzinou, A., and Fablet,
  R. (2025).
\newblock Neural prediction of lagrangian drift trajectories on the sea
  surface.
\newblock {\em Artificial Intelligence for the Earth Systems}.

\bibitem[Breivik et~al., 2013]{search_and_rescue}
Breivik, O., Allen, A.~A., Maisondieu, C., and Olagnon, M. (2013).
\newblock Advances in search and rescue at sea.
\newblock {\em Ocean Dynamics}, 63:83--88.
\newblock Publisher: Springer.

\bibitem[Cancet et~al., 2019]{cancet2019evaluation}
Cancet, M., Griffin, D., Cahill, M., Chapron, B., Johannessen, J., and Donlon,
  C. (2019).
\newblock Evaluation of globcurrent surface ocean current products: A case
  study in australia.
\newblock {\em Remote sensing of environment}, 220:71--93.

\bibitem[Checkley and Barth, 2009]{pacific_dynamics}
Checkley, D.~M. and Barth, J.~A. (2009).
\newblock Patterns and processes in the california current system.
\newblock {\em Progress in Oceanography}, 83(1-4):49--64.

\bibitem[Chelton et~al., 2011]{chelton2011global}
Chelton, D.~B., Schlax, M.~G., and Samelson, R.~M. (2011).
\newblock Global observations of nonlinear mesoscale eddies.
\newblock {\em Progress in oceanography}, 91(2):167--216.

\bibitem[Dagestad and R{\"o}hrs, 2019]{dagestad2019prediction}
Dagestad, K.-F. and R{\"o}hrs, J. (2019).
\newblock Prediction of ocean surface trajectories using satellite derived vs.
  modeled ocean currents.
\newblock {\em Remote sensing of environment}, 223:130--142.

\bibitem[de~Pablo et~al., 2020]{de2020mohid}
de~Pablo, H., Garaboa-Paz, D., Canelas, R., Campuzano, F., and Neves, R.
  (2020).
\newblock Mohid-lagrangian: A lagrangian transport model from local to globals
  scales. applications to the marine litter problem.
\newblock In {\em EGU General Assembly Conference Abstracts}, page 21895.

\bibitem[Dee et~al., 2011]{dee2011era}
Dee, D.~P., Uppala, S.~M., Simmons, A.~J., Berrisford, P., Poli, P., Kobayashi,
  S., Andrae, U., Balmaseda, M.~A., Balsamo, G., and Bauer, P. e.~a. (2011).
\newblock The {ERA}-interim reanalysis: Configuration and performance of the
  data assimilation system.
\newblock {\em Quarterly Journal of the royal meteorological society},
  137(656):553--597.
\newblock Publisher: Wiley Online Library.

\bibitem[Della~Cioppa and Buongiorno~Nardelli, 2025]{della2025predicting}
Della~Cioppa, L. and Buongiorno~Nardelli, B. (2025).
\newblock Predicting oceanic lagrangian trajectories with hybrid space-time cnn
  architecture.
\newblock {\em EGUsphere}, 2025:1--18.

\bibitem[Dewar and Bane, 1989]{dewar1989gulf}
Dewar, W.~K. and Bane, J.~M. (1989).
\newblock Gulf stream dynamics. pad {II}: Eddy energetics at 73 w.
\newblock {\em Journal of Physical Oceanography}, 19(10):1574--1587.

\bibitem[Dufau et~al., 2016]{dufau2016mesoscale}
Dufau, C., Orsztynowicz, M., Dibarboure, G., Morrow, R., and Le~Traon, P.-Y.
  (2016).
\newblock Mesoscale resolution capability of altimetry: Present and future.
\newblock {\em Journal of Geophysical Research: Oceans}, 121(7):4910--4927.
\newblock Publisher: Wiley Online Library.

\bibitem[Etienne et~al., 2023]{etienneglobal}
Etienne, H., Verbrugge, N., Boone, C., Rubio, A., Solabarrieta, L., Corgnati,
  L., Mantovani, C., Reyes, E., Chifflet, M., and Mader, J. e.~a. (2023).
\newblock Quality information document: Global ocean-delayed mode in-situ
  observations of surface (drifters and hfr) and sub-surface (vessel-mounted
  adcps) water velocity.

\bibitem[Fablet et~al., 2023]{fablet2023multimodal}
Fablet, R., Febvre, Q., and Chapron, B. (2023).
\newblock Multimodal 4dvarnets for the reconstruction of sea surface dynamics
  from {SST}-{SSH} synergies.
\newblock {\em {IEEE} Transactions on Geoscience and Remote Sensing}.
\newblock Publisher: {IEEE}.

\bibitem[Fajardo-Urbina et~al., 2024]{fajardo2024efficient}
Fajardo-Urbina, J.~M., Liu, Y., Georgievska, S., Gr{\"a}we, U., Clercx, H.~J.,
  Gerkema, T., and Duran-Matute, M. (2024).
\newblock Efficient deep learning surrogate method for predicting the transport
  of particle patches in coastal environments.
\newblock {\em Marine Pollution Bulletin}, 209:117251.

\bibitem[Fu and Ubelmann, 2014]{fu2014transition}
Fu, L.-L. and Ubelmann, C. (2014).
\newblock On the transition from profile altimeter to swath altimeter for
  observing global ocean surface topography.
\newblock {\em Journal of Atmospheric and Oceanic Technology}, 31(2):560--568.

\bibitem[Good et~al., 2020]{good2020current}
Good, S., Fiedler, E., Mao, C., Martin, M.~J., Maycock, A., Reid, R.,
  Roberts-Jones, J., Searle, T., Waters, J., While, J., et~al. (2020).
\newblock The current configuration of the ostia system for operational
  production of foundation sea surface temperature and ice concentration
  analyses.
\newblock {\em Remote Sensing}, 12(4):720.

\bibitem[Gula et~al., 2015]{gula2015gulf}
Gula, J., Molemaker, M.~J., and McWilliams, J.~C. (2015).
\newblock Gulf stream dynamics along the southeastern {US} seaboard.
\newblock {\em Journal of Physical Oceanography}, 45(3):690--715.
\newblock Publisher: American Meteorological Society.

\bibitem[Jenkins et~al., 2023]{jenkins2023dnn}
Jenkins, J., Paiement, A., Ourmières, Y., Le~Sommer, J., Verron, J., Ubelmann,
  C., and Glotin, H. (2023).
\newblock A {DNN} framework for learning lagrangian drift with uncertainty.
\newblock {\em Applied Intelligence}, 53(20):23729--23739.
\newblock Publisher: Springer.

\bibitem[Kang et~al., 2005]{kang2005scale}
Kang, Y., Morooka, K., and Nagahashi, H. (2005).
\newblock Scale invariant texture analysis using multi-scale local
  autocorrelation features.
\newblock In {\em Scale Space and PDE Methods in Computer Vision: 5th
  International Conference, Scale-Space 2005, Hofgeismar, Germany, April 7-9,
  2005. Proceedings 5}, pages 363--373. Springer.

\bibitem[Krauß and Böning, 1987]{lagr_time_scale}
Krauß, W. and Böning, C.~W. (1987).
\newblock Lagrangian properties of eddy fields in the northern north atlantic
  as deduced from satellite-tracked buoys.
\newblock {\em Journal of Marine Research}, 45(2):259--291.
\newblock Publisher: Sears Foundation for Marine Research.

\bibitem[Lange and van Sebille, 2017]{Lange2017}
Lange, M. and van Sebille, E. (2017).
\newblock Parcels: Efficient lagrangian particle tracking software with a
  user-friendly interface.
\newblock {\em Environmental Modelling \& Software}, 49:53--60.

\bibitem[Le~Guillou et~al., 2025]{le2025vardyn}
Le~Guillou, F., Chapron, B., and Rio, M.-H. (2025).
\newblock Vardyn: Dynamical joint-reconstructions of sea surface height and
  temperature from multi-sensor satellite observations.
\newblock {\em Journal of Advances in Modeling Earth Systems},
  17(4):e2024MS004689.

\bibitem[Le~Guillou et~al., 2021]{le2021joint}
Le~Guillou, F., Lahaye, N., Ubelmann, C., Metref, S., Cosme, E., Ponte, A.,
  Le~Sommer, J., Blayo, E., and Vidard, A. (2021).
\newblock Joint estimation of balanced motions and internal tides from future
  wide-swath altimetry.
\newblock {\em Journal of Advances in Modeling Earth Systems},
  13(12):e2021MS002613.

\bibitem[Le~Traon et~al., 2019]{le2019observation}
Le~Traon, P.~Y., Reppucci, A., Alvarez~Fanjul, E., Aouf, L., Behrens, A.,
  Belmonte, M., Bentamy, A., Bertino, L., Brando, V.~E., Kreiner, M.~B., et~al.
  (2019).
\newblock From observation to information and users: The copernicus marine
  service perspective.
\newblock {\em Frontiers in marine science}, 6:234.

\bibitem[Lellouche et~al., 2021]{glo12}
Lellouche, J.~M., Greiner, E., Bourdalle-Badie, R., Garric, G., Melet, A.,
  Drévillon, M., Bricaud, C., and et~al. (2021).
\newblock The copernicus global 1/12 oceanic and sea ice {GLORYS}12 reanalysis.
\newblock {\em Frontiers in Earth Science}, page 698876.
\newblock Publisher: Frontiers Media {SA}.

\bibitem[Lellouche et~al., 2018]{lellouche_glorys12}
Lellouche, J.~M., Greiner, E., Le~Galloudec, O., Garric, G., Regnier, C.,
  Drevillon, M., Benkiran, M., Testut, C.-E., Bourdalle-Badie, R., and
  Gasparin, F. e.~a. (2018).
\newblock Recent upyears to the copernicus marine service global ocean
  monitoring and forecasting real-time 1/ 12 high-resolution system.
\newblock {\em Ocean Science}, 14(5):1093--1126.
\newblock Publisher: Copernicus {GmbH}.

\bibitem[Liu and Wang, 2025]{liu2025detection}
Liu, X. and Wang, M. (2025).
\newblock Detection of ocean eddies from satellite ocean color and sst
  measurements using a deep learning approach.
\newblock {\em International Journal of Applied Earth Observation and
  Geoinformation}, 144:104929.

\bibitem[Liu and Weisberg, 2011]{liu2011evaluation}
Liu, Y. and Weisberg, R.~H. (2011).
\newblock Evaluation of trajectory modeling in different dynamic regions using
  normalized cumulative lagrangian separation.
\newblock {\em Journal of Geophysical Research: Oceans}, 116(C9).
\newblock Publisher: Wiley Online Library.

\bibitem[Liu et~al., 2011]{Liu2011Tracking}
Liu, Y., WeisBerg, R.~H., Hu, C., and Zheng, L. (2011).
\newblock Tracking the deepwater horizon oil spill: A modeling perspective.
\newblock {\em Eos, Trans. Amer. Geophys. Union}, 92:45--46.

\bibitem[Liu et~al., 2014]{liu2014evaluation}
Liu, Y., Weisberg, R.~H., Vignudelli, S., and Mitchum, G.~T. (2014).
\newblock Evaluation of altimetry-derived surface current products using
  lagrangian drifter trajectories in the eastern gulf of mexico.
\newblock {\em Journal of Geophysical Research: Oceans}, 119(5):2827--2842.

\bibitem[Lumpkin and Elipot, 2010]{lumpkin2010surface}
Lumpkin, R. and Elipot, S. (2010).
\newblock Surface drifter pair spreading in the north atlantic.
\newblock {\em Journal of Geophysical Research: Oceans}, 115(C12).

\bibitem[Martin et~al., 2023]{martin2023synthesizing}
Martin, S.~A., Manucharyan, G.~E., and Klein, P. (2023).
\newblock Synthesizing sea surface temperature and satellite altimetry
  observations using deep learning improves the accuracy and resolution of
  gridded sea surface height anomalies.
\newblock {\em Journal of Advances in Modeling Earth Systems},
  15(5):e2022MS003589.

\bibitem[Morrow and Le~Traon, 2012]{morrow2012recent}
Morrow, R. and Le~Traon, P.-Y. (2012).
\newblock Recent advances in observing mesoscale ocean dynamics with satellite
  altimetry.
\newblock {\em Advances in Space Research}, 50(8):1062--1076.

\bibitem[Niiler and Paduan, 1995]{niiler1995wind}
Niiler, P.~P. and Paduan, J.~D. (1995).
\newblock Wind-driven motions in the northeast pacific as measured by
  lagrangian drifters.
\newblock {\em Journal of Physical Oceanography}, 25(11):2819--2830.

\bibitem[Onink et~al., 2019]{onink2019role}
Onink, V., Wichmann, D., Delandmeter, P., and van Sebille, E. (2019).
\newblock The role of ekman currents, geostrophy, and stokes drift in the
  accumulation of floating microplastic.
\newblock {\em Journal of Geophysical Research: Oceans}, 124(3):1474--1490.

\bibitem[{\"O}zg{\"o}kmen et~al., 2000]{ozgokmen2000predictability}
{\"O}zg{\"o}kmen, T.~M., Griffa, A., Mariano, A.~J., and Piterbarg, L.~I.
  (2000).
\newblock On the predictability of lagrangian trajectories in the ocean.
\newblock {\em Journal of Atmospheric and Oceanic Technology}, 17(3):366--383.

\bibitem[P{\"a}rn et~al., 2023]{parn2023effects}
P{\"a}rn, O., Davulien{\.e}, L., Moy, D.~M., Vahter, K., Stips, A., and
  Torsvik, T. (2023).
\newblock Effects of eulerian current, stokes drift and wind while simulating
  surface drifter trajectories in the baltic sea.
\newblock {\em Oceanologia}, 65(3):453--465.

\bibitem[Pawar et~al., 2016]{pawar2016plastic}
Pawar, P.~R., Shirgaonkar, S.~S., and Patil, R.~B. (2016).
\newblock Plastic marine debris: Sources, distribution and impacts on coastal
  and ocean biodiversity.
\newblock {\em PENCIL Publication of Biological Sciences}, 3(1):40--54.

\bibitem[Poulain et~al., 2009]{poulain2009wind}
Poulain, P.-M., Gerin, R., Mauri, E., and Pennel, R. (2009).
\newblock Wind effects on drogued and undrogued drifters in the eastern
  mediterranean.
\newblock {\em Journal of Atmospheric and Oceanic Technology},
  26(6):1144--1156.

\bibitem[Pujol et~al., 2016]{pujol2016duacs}
Pujol, M.-I., Faugère, Y., Taburet, G., Dupuy, S., Pelloquin, C., Ablain, M.,
  and Picot, N. (2016).
\newblock {DUACS} {DT}2014: the new multi-mission altimeter data set
  reprocessed over 20 years.
\newblock {\em Ocean Science}, 12(5):1067--1090.
\newblock Publisher: Copernicus {GmbH}.

\bibitem[Ralph and Niiler, 1999]{ralph1999wind}
Ralph, E.~A. and Niiler, P.~P. (1999).
\newblock Wind-driven currents in the tropical pacific.
\newblock {\em Journal of Physical Oceanography}, 29(9):2121--2129.

\bibitem[Rio et~al., 2014]{rio2014beyond}
Rio, M.-H., Mulet, S., and Picot, N. (2014).
\newblock Beyond goce for the ocean circulation estimate: Synergetic use of
  altimetry, gravimetry, and in situ data provides new insight into geostrophic
  and ekman currents.
\newblock {\em Geophysical Research Letters}, 41(24):8918--8925.

\bibitem[Scott et~al., 2012]{scott2012estimates}
Scott, R.~B., Ferry, N., Drévillon, M., Barron, C.~N., Jourdain, N.~C., and
  Lellouche, J. M. e.~a. (2012).
\newblock Estimates of surface drifter trajectories in the equatorial atlantic:
  a multi-model ensemble approach.
\newblock {\em Ocean Dynamics}, 62:1091--1109.
\newblock Publisher: Springer.

\bibitem[Seo et~al., 2016]{seo2016eddy}
Seo, H., Miller, A.~J., and Norris, J.~R. (2016).
\newblock Eddy--wind interaction in the california current system: Dynamics and
  impacts.
\newblock {\em Journal of Physical Oceanography}, 46(2):439--459.

\bibitem[Taburet et~al., 2019]{taburet2019duacs}
Taburet, G., Sanchez-Roman, A., Ballarotta, M., Pujol, M.-I., Legeais, J.-F.,
  Fournier, F., Faugere, Y., and Dibarboure, G. (2019).
\newblock Duacs dt2018: 25 years of reprocessed sea level altimetry products.
\newblock {\em Ocean Science}, 15(5):1207--1224.

\bibitem[Tang et~al., 2022]{tang2022mesoscale}
Tang, R., Yu, Y., Xi, J., Ma, W., and Wang, Y. (2022).
\newblock Mesoscale eddies induce variability in the sea surface temperature
  gradient in the kuroshio extension.
\newblock {\em Frontiers in Marine Science}, 9:926954.

\bibitem[Trong et~al., 2025]{trong2025comparative}
Trong, N., Thanh, P., Quang, P., Tinh, L., Manh, V., Vinh, T., Elshewy, M.,
  et~al. (2025).
\newblock Comparative analysis of prediction accuracy for drifting buoy data
  using cnn (conv1d) and gru deep learning models with varying data volumes.
\newblock {\em International Journal of Geoinformatics}, 21(4):115--130.

\bibitem[Van~Sebille et~al., 2020]{van2020physical}
Van~Sebille, E., Aliani, S., Law, K.~L., Maximenko, N., Alsina, J.~M., Bagaev,
  A., Bergmann, M., Chapron, B., Chubarenko, I., C{\'o}zar, A., et~al. (2020).
\newblock The physical oceanography of the transport of floating marine debris.
\newblock {\em Environmental Research Letters}, 15(2):023003.

\bibitem[Verrier et~al., 2017]{verrier2017assessing}
Verrier, S., Le~Traon, P.-Y., and Remy, E. (2017).
\newblock Assessing the impact of multiple altimeter missions and argo in a
  global eddy-permitting data assimilation system.
\newblock {\em Ocean Science}, 13(6):1077--1092.

\bibitem[Verrier et~al., 2018]{verrier2018assessing}
Verrier, S., Le~Traon, P.-Y., Remy, E., and Lellouche, J.~M. (2018).
\newblock Assessing the impact of {SAR} altimetry for global ocean analysis and
  forecasting.
\newblock {\em Journal of Operational Oceanography}, 11(2):82--86.
\newblock Publisher: Taylor \& Francis.

\bibitem[Visser, 2008]{lagrangian_fokker_planck_plankton}
Visser, A.~W. (2008).
\newblock Lagrangian modelling of plankton motion: From deceptively simple
  random walks to fokker–planck and back again.
\newblock {\em Journal of Marine Systems}, 70(3-4):287--299.
\newblock Publisher: Elsevier.

\bibitem[Wunsch, 1999]{wunsch1999interpretation}
Wunsch, C. (1999).
\newblock The interpretation of short climate records, with comments on the
  north atlantic and southern oscillations.
\newblock {\em Bulletin of the american meteorological society},
  80(2):245--256.

\bibitem[Zhang et~al., 2020]{zhang2020evaluation}
Zhang, X., Cheng, L., Zhang, F., Wu, J., Li, S., Liu, J., Chu, S., Xia, N.,
  Min, K., Zuo, X., et~al. (2020).
\newblock Evaluation of multi-source forcing datasets for drift trajectory
  prediction using lagrangian models in the south china sea.
\newblock {\em Applied Ocean Research}, 104:102395.

\end{thebibliography}

\end{document}